\documentclass[twocolumn,superscriptaddress,preprintnumbers,amsmath,amssymb,floatfix,nofootinbib]{revtex4-1}

\usepackage{amsmath}
\usepackage{amssymb}
\usepackage{graphicx}
\usepackage{color}
\usepackage{graphicx}
\usepackage{upgreek}
\usepackage{scalerel}
\usepackage{multirow}
\usepackage{pgffor}
\usepackage{pdfpages}
\usepackage{ mathdots}
\usepackage{float}
\usepackage{comment}  
\makeatletter
\AtBeginDocument{\let\LS@rot\@undefined}
\makeatother

\begin{document}

\title{Quantifying the information impact of future searches for exoplanetary biosignatures}

\author{Amedeo Balbi}\email{amedeo.balbi@roma2.infn.it}
\affiliation{Dipartimento di Fisica, Universit\`a di Roma ``Tor Vergata'', Via della Ricerca Scientifica, 00133 Roma, Italy}
\author{Claudio Grimaldi}\email{claudio.grimaldi@epfl.ch}
\affiliation{Laboratory of Physics of Complex Matter, Ecole Polytechnique F\'ed\'erale
de Lausanne, Station 3, CP-1015 Lausanne, Switzerland} 
\affiliation{Centro Studi e Ricerche Enrico Fermi, Piazza del Viminale, 1, 00184 Roma, Italy}

\begin{abstract}
One of the major goals for astronomy in the next decades is the remote search for biosignatures (i.e.\ the spectroscopic evidence of biological activity) in exoplanets. Here, we adopt a Bayesian statistical framework to discuss the implications of such future searches, both in the case when life is detected, and when no definite evidence is found. 
We show that even a single detection of biosignatures in the vicinity of our stellar system, in a survey of similar size to what will be obtainable in the next two decades, would affect significantly our prior belief on the frequency of life in the universe, even starting from a neutral or pessimistic stance. In particular, after such discovery, an initially agnostic observer would be led to conclude that there are more than $10^5$ inhabited planets in the galaxy with a probability exceeding $95$\%. However, this conclusion would be somewhat weakened by the viability of transfer of biological material over interstellar distances, as in panspermia scenarios.
Conversely, the lack of significant evidence of biosignatures would have little effect, leaving the assessment of the abundance of life in the galaxy still largely undetermined. 
\end{abstract}

\maketitle

Over the past $2$ decades, astronomical observations have detected thousands of planets orbiting other stars in our galaxy, allowing to draw robust statistical conclusions on the populations of such planets\cite{Batalha2014}. Generally speaking, it is now believed that every star in our galaxy should have at least one planet\cite{Cassan2012}, and that many such planets have physical features that may be conducive to the presence of life\cite{Petigura2013, Zink2019, Dressing2015}.

With the focus of current research rapidly shifting from the detection of exoplanets to their characterization -- and, in particular, to the study of their atmospheric composition -- we are getting closer to the goal of looking for spectroscopic signatures of biological activity on other worlds\cite{Seager2014,Grenfell2017,Fujii2018,Madhusudhan2019}. In the near term, the TESS\cite{Ricker2014}, CHEOPS\cite{Broeg2018} and PLATO\cite{Rauer2014} space missions will refine the sample of potentially habitable nearby planets more suitable for follow-up observations. Over the next couple of decades there will be realistic opportunities for attempting the detection of biosignatures on the most promising targets, both from the ground (e.g.\ with the European Extremely Large Telescope\footnote{https://www.eso.org/sci/facilities/eelt/
}) and with dedicated space observatories (such as JWST\cite{Gardner2006} or ARIEL\cite{Tinetti2018}). On a longer time scale, envisioned missions such Habitable Exoplanet Observatory (HabEX\footnote{http://www.jpl.nasa.gov/habex/}), the Large UV/Optical/IR Surveyor (LUVOIR\footnote{https://asd.gsfc.nasa.gov/luvoir/}) and the Origins Space Telescope\footnote{https://asd.gsfc.nasa.gov/firs/} might attempt biosignature detection through the direct imaging of habitable rocky exoplanets.

Since technological limitations will initially restrict the search for biosignatures to the immediate vicinity of our stellar system (i.e.\ within a few tens light years), a rigorous statistical treatment will be necessary in order to draw conclusions on the possible distribution of inhabited planets in the entire galaxy from a survey of limited spatial extent. This will be true both in the case of a positive detection of life on one or more exoplanets in a given volume, and in the case where no evidence will be found. 

Here, we suggest an approach to this problem based on the adoption of a Bayesian perspective, showing how existing knowledge or credence on the presence of life beyond Earth will be updated as new evidence will be collected from future missions. A notable previous application of the Bayesian methodology
in the context of life emerging in the universe was the attempt to quantify the rate of abiogenesis conditioned on a single datum, i.e.\ the early appearance of life on Earth, combined with the evidence that it took $\approx 3.8$ Gyr for life to evolve intelligence\cite{Spiegel2012}. Further developments along this line considered how future evidence would update our previous knowledge on the rate of abiogenesis\cite{Chen2018}. 

Our study tackles the issue of how frequent life is in the universe from a different perspective. We bypass the question of the timescales involved in the abiogenesis and we rather focus on the present abundance of inhabited planets in the galaxy. In particular, we are interested in assessing the impact of new data (those that could be possibly collected in the next $2$ decades) in terms of information gain with respect to existing credence on the probability of life on other planets. We suggest a way to disentangle this unknown probability from others that can be in principle estimated independently, in particular, those pertaining to the probability that a specific survey can in fact observe habitable planets. A related relevant question addressed by our study is how hypotheses that assign a lower or higher credence to the presence of life outside Earth---i.e.\ a pessimistic, neutral or optimistic attitude towards extraterrestrial life---are weighed and compared in light of new, sparse evidence. Finally, we consider how our results are altered when accounting for the possibility that the distribution of life is correlated over some characteristic distance, such as in panspermia scenarios.

\section*{Methods}
\subsection*{Main assumptions}

Our statistical model assumes that there are $N$ potentially habitable planets in the Milky Way (i.e.\ rocky planets orbiting the habitable zone of their host star), and that a survey has looked for spectroscopic biosignatures within a radius $R$ centered around Earth.  Statistical estimates based on available data suggest that the fraction of Sun-like (GK-type) and M dwarf stars in our galaxy hosting rocky planets in the habitable zone is about 10\%-20\% and 24\% respectively \cite{Petigura2013,Zink2019,Dressing2015}, resulting in a number of potentially habitable planets of order $N=10^{10}$.  We adopt this estimate as a fiducial value for $N$ in our analysis, without referring to the particular spectral type of the host star (for a recent analysis of this issue see \cite{Haqq-Misra2018}). We further assume that the probability of detecting biosignatures within the survey volume is $p$, so that the expected number of biosignature detections is 
\begin{equation}
\label{bark1}
\bar{k}(R)=pN\pi(R)
\end{equation}
where $\pi(R)$ is the probability of a habitable planet being within $R$, given by:
\begin{equation}
\label{pi0}
\pi(R)=N^{-1}\int\! d\mathbf{r}\rho(\mathbf{r})\theta(R-\vert\mathbf{r}-\mathbf{r}_\textrm{E}\vert),
\end{equation}
where $\mathbf{r}_\textrm{E}$ is the position vector of the Earth relative to the galactic center, and $\theta(x)$ is the Heavyside step function. The number density function, $\rho(\mathbf{r})$, is defined in such a way that $\rho(\mathbf{r})d\mathbf{r}$ gives the expected number of habitable planets within the volume element $d\mathbf{r}$ about $\mathbf{r}$, so that $\int\! d\mathbf{r}\rho(\mathbf{r})=N$.

The probability $p$ is a shorthand for the various factors that concur to make the presence of detectable biosignatures possible. In our Bayesian analysis we distinguish the factors ascribed to the selection effects of a specific survey from those that are truly inherent to the presence of biosignatures. To this end, we adopt a formalism similar to the one first suggested in \cite{Seager2018}: this is akin to the Drake equation\cite{Drake1965} used in the context of the search for extraterrestrial intelligence, but adapted to the search for biosignatures. In our notation, this reduces to writing down the probability $p$ as the product of independent probabilities:
\begin{equation}
\label{prob_p}
p=p_{\rm a}p_{\rm d}p_{\rm l}.
\end{equation}

The first probability, $p_{\rm a}$, pertains to astrophysical factors and observational limitations. Given a exoplanetary survey, only a fraction of systems will be suitable for the search of biosignatures. For example, one may look only for planets in the habitable zone of specific types of stars. The value of $p_a$ can also account for the fact that not all planets in the habitable zone of their star will indeed be habitable. Furthermore, there are other selection effects involved in the specific observational strategy: for example, in a transit survey there will be strict requirements on the geometrical configuration of the orbital plane, while a direct imaging survey will be limited by the variability of the reflected starlight as the planet orbits the star. In principle, a good estimate of $p_{\rm a}$ can be obtained from astrophysical and observational considerations. Eventually, a given survey will only sample the quantity $N \pi (R) p_{\rm a}$. For example, the number of planets that can be scanned for biosignatures following the TESS survey can be estimated to be $\approx 4$, while it would be $\approx 11$ for future ground-based imaging\cite{Seager2018}. 

The other probability factors in \eqref{prob_p}, $p_{\rm d}$ and $p_{\rm l}$, are not related to a specific survey and pertain exclusively to the likelihood that life-harboring planets in the galaxy display biosignatures.
The probability $p_{\rm d}$ quantifies the fact that, in general, detectable biosignatures are not expected to accompany all instances of life on a planet. For example, chemical byproducts of life can significantly alter an exoplanet atmosphere only after some time has passed from the appearance of life. Furthermore, depending on geological and astrophysical factors, life might go extinct after a few hundred million years, as it may have happened on Mars. If we focus on free molecular oxygen as the quintessential biosignature gas, this has been remotely detectable in the Earth atmosphere for $\approx 2$ Gyr, roughly half the Earth's age. If we take this as representative of the average, this would point to $p_{\rm d}\approx 0.5$, i.e.\ a large probability. Of course, there is no telling if this is a universal feature of any biosphere, but it nevertheless hints to a significant upper limit on $p_{\rm d}$. 

Finally, $p_{\rm l}$ is the probability that life indeed appears on a habitable planet besides Earth. This is, essentially, the probability of abiogenesis, and is a truly unknown factor, which makes $\bar{k}=p_{\rm d}p_{\rm l}N$, the expected number of planets with biosignatures in the Milky Way, highly indeterminate.

In the present work, we left unaddressed the possibility of both false negatives (biosignatures that are present but go undetected) and false positives (gases of abiotic origin that are mistakenly interpreted as products of life): however, we note that in principle both can be incorporated in our formalism through another probability factor, following, for example, the Bayesian framework outlined in \cite{Walker2018, Catling2018} (\textit{SI Appendix}, section II). Our procedure could also be easily specialized for technosignatures, incorporating the appropriate probabilistic factors, along the lines of \cite{Lingam2019}.

In modeling $\pi(R)$ we focus on the thin disk component of the Galaxy and adopt an axisymmetric model of the number density of exoplanets:
\begin{equation}
\label{ghz}
\rho(\mathbf{r})=N\frac{ e^{-\displaystyle r/r_s}e^{-\displaystyle \vert z\vert/z_s}}{4\pi r_s^2 z_s},
\end{equation}
where $r$ is the radial distance from the galactic center, $z$ is the height from the galactic plane, $r_s=8.15$ kly, and $z_s=0.52$ kly \cite{Misiriotis2006}.
For $R$ smaller than about $1$ kly and taking $r_\textrm{E}\simeq 27$ kly, the Taylor expansion of $\pi(R)$ for small $R$
yields $\pi(R)\simeq (4\pi/3)\rho(\mathbf{r}_\textrm{E})R^3/N=(R/a)^3$, with $a=14.2$ kly. Although \eqref{ghz} assumes that the
density profile of habitable exoplanets is proportional to that of stars in the Galaxy, other factors such as the metallicity gradient may
affect the overall radial dependence of $\rho(\mathbf{r})$ (\textit{SI Appendix}, section III).

Throughout this work, we take an observational radius of $R=100$ ly which, although corresponding to a galactic fractional volume of only  $\pi(R=100 \textrm{\ ly})\simeq 3.5\times 10^{-7}$, is an optimistic upper limit of the search range attainable over the next couple of decades. In choosing $p_{\rm a}$ we consider two limiting situations: i) $p_{\rm a} = 1$, which corresponds to an ideal survey that has searched for biosignatures in all the existing habitable planets within a given distance $R$ from Earth; ii) $p_{\rm a} \rightarrow 0$, which corresponds to an exceedingly small number of targeted planets compared to initial sample size (\textit{SI Appendix}, section I).

The probability that a survey searching for biosignature within $R$ finds remotely detectable biosignatures on exactly $k=0,\, 1,\, 2,\, \ldots$ exoplanets 
follows a binomial distribution:
\begin{equation}
\label{binom2}
P_k(R)=\binom{N}{k} [\pi(R)p]^k [1-\pi(R)p]^{N-k},
\end{equation}
The average number of exoplanets detectable by the survey is 
 $\bar{k}(R)=N p \pi(R)$, so that by keeping $\bar{k}(R)$ finite, the large $N$ limit of $P_k(R)$ reduces to a Poissonian distribution:
\begin{equation}
\label{poisson1}
P_k(R)=\frac{[\bar{k}(R)]^k}{k!}e^{-\bar{k}(R)}.
\end{equation}

By rewriting \eqref{bark1} as: 
	\begin{equation}
	\label{kbar2}
	\bar{k}(R)=\bar{k}p_{\rm a}\pi(R),
	\end{equation} 
	our analysis translates the outcome of a search for biosignatures into an increase in the posterior information on $\bar{k}$. In practice, we use Bayes theorem to update the prior probability distribution function (PDF) of $\bar{k}$, after gathering the evidence that exactly $k$ biosignatures are detected in the survey, which is parametrized by $p_{\rm a}\pi(R)$.	

\subsection*{Bayesian analysis}

By isolating the probability factor in $p$ that pertains to astrophysical and observational constraints, $p_{\rm a}$, from those referring to the probability
of abiogenesis and formation of biotic atmospheres, $p_{\rm d}$ and $p_{\rm l}$, we parametrize the survey by $p_{\rm a}\pi(R)$ and the
expected number of exoplanets in the entire galaxy producing biosignatures by $\bar{k}=p_{\rm d}p_{\rm l}N$. Next, we denote $\mathcal{E}_k$
the event of detecting exactly $k$ biosignatures during the survey, so that using $\bar{k}(R)=\bar{k}p_{\rm a}\pi(R)$ \eqref{poisson1} gives
the likelihood of $\mathcal{E}_k$ being true given $\bar{k}$:
\begin{equation}
\label{poisson2}
P(\mathcal{E}_k\vert \bar{k})=\frac{[\bar{k}p_{\rm a}\pi(R)]^k}{k!}e^{-\bar{k}p_{\rm a}\pi(R)}.
\end{equation} 
We aim to find the posterior PDF of $\bar{k}$ resulting from the event $\mathcal{E}_k$. To this end we consider 
the prior PDF of $\bar{k}$, that is, the probability distribution we ascribe to $\bar{k}$ before gathering the evidence $\mathcal{E}_k$. In the following,
we will refer to a specific functional form of the prior PDF as a model $M$: $p(\bar{k}\vert M)$. Following the logic of Bayes' theorem, the
posterior PDF is thus obtained from:
\begin{equation}
\label{bayes1}
p(\bar{k}\vert \mathcal{E}_k, M)=\frac{P(\mathcal{E}_k\vert \bar{k})p(\bar{k}\vert M)}{P(\mathcal{E}_k\vert M)},
\end{equation}
where 
\begin{equation}
\label{bayesM3}
P(\mathcal{E}_k\vert M_i)=\int\!d\bar{k} P(\mathcal{E}_k\vert \bar{k})p(\bar{k}\vert M_i)
\end{equation}
is the likelihood of $\mathcal{E}_k$ given the model $M$. 

We consider three different models of the prior defined in the interval $\bar{k}_\textrm{min}$ to $\bar{k}_\textrm{max}$ labelled by the subscript $i=0,\ 1,\ 2$:
\begin{equation}
\label{priormodels}
p(\bar{k}\vert M_i)\propto\bar{k}^{-i},\,\,\,\textrm{for}\,\,\bar{k}_\textrm{min}\leq\bar{k}\leq\bar{k}_\textrm{max},
\end{equation} 
where $i=0$ gives a prior PDF uniform in $\bar{k}$, which strongly favors large values of $\bar{k}$ (optimistic model $M_0$), $i=1$ corresponds to a 
non-informed prior which is log-uniformly distributed in the interval $\bar{k}_\textrm{min}$ to $\bar{k}_\textrm{max}$ (non-informed model $M_1$), 
and $i=2$ gives a highly informative prior favoring small values of $\bar{k}$ (pessimistic model $M_2$).

Finally, we consider here
only two events resulting from the survey: non-detection, $\mathcal{E}_0$, and detection of one biosignature, $\mathcal{E}_1$, (\textit{SI Appendix}, section I).

\begin{figure*}[t]
	\begin{center}
		\includegraphics[width=0.85\textwidth,clip=true]{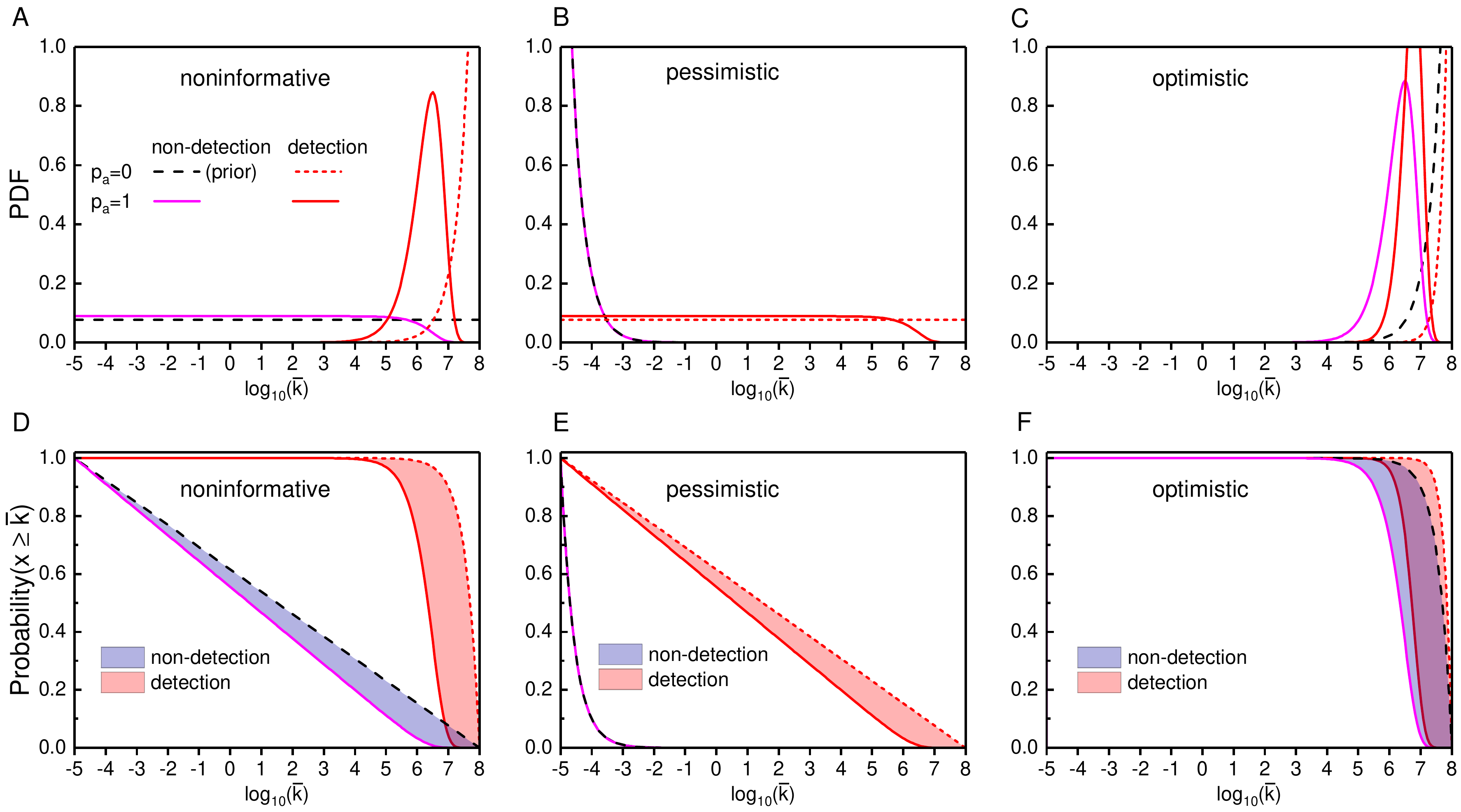}
		\caption{Results for a survey searching for atmospheric biosignatures within a distance $R=100$ ly from Earth. Shown in each row, from left to right, are the posterior PDF (upper row) and CCDF (bottom row), updated in light of the evidence, starting from a non-informative, pessimistic and optimistic prior (black dashed curves), respectively. The continuous curves refer to the posterior PDF and CCDF for the case $p_a=1$ (all habitable planets in the survey observed). The limit $p_{\rm a}=0$ is shown by red short-dashed curves in the case of detection, while the posteriors resulting from non-detection at $p_{\rm a}=0$ coincide with the priors. The shaded areas in the CCDF encompass the limiting cases $p_a=0$ and $p_a=1$, giving the range of probabilities that the mean number of life-bearing planets is larger than $\bar{k}$.
		}\label{fig1}
	\end{center}
\end{figure*}

\section*{Results and discussion}
\subsection*{Non-informative prior}

We start by assuming no prior knowledge on even the scale of $\bar{k}$: this is modeled by taking the non-informed log-uniform prior $p(\bar{k}\vert M_1)$,
which gives equal weight to all orders of magnitude of $\bar{k}$. We take initially $\bar{k}_\textrm{max}=10^{-2}N=10^8$, which corresponds to assuming that at most one planet out of $100$ has detectable biosignatures. In making a choice for $\bar{k}_\textrm{min}$ one may be tempted to take $\bar{k}_\textrm{min}=1$ because we know for sure that at least one planet in the Milky Way (the Earth) harbors life. This choice would be justifiable if we were interested in calculating the posterior PDF of $\bar{k}$ from the evidences gathered within a distance $R$ from a randomly chosen point in the Galaxy. However, with the exclusion of the Earth, we ignore whether other planets harbor life (either detectable or not). From our standpoint, therefore, $\bar{k}_\textrm{min}$ can be well below $1$. Here we take for the sake of illustration $\bar{k}_\textrm{min}=10^{-5}$: to give an idea of how small this is, it corresponds to having roughly just one planet with biosignatures in $10^5$ hypothetical random realizations of the Milky Way galaxy (we investigate the effect of varying $\bar{k}_\textrm{min}$ below).

\begin{figure*}[t]
	\begin{center}
		\includegraphics[width=0.7\textwidth,clip=true]{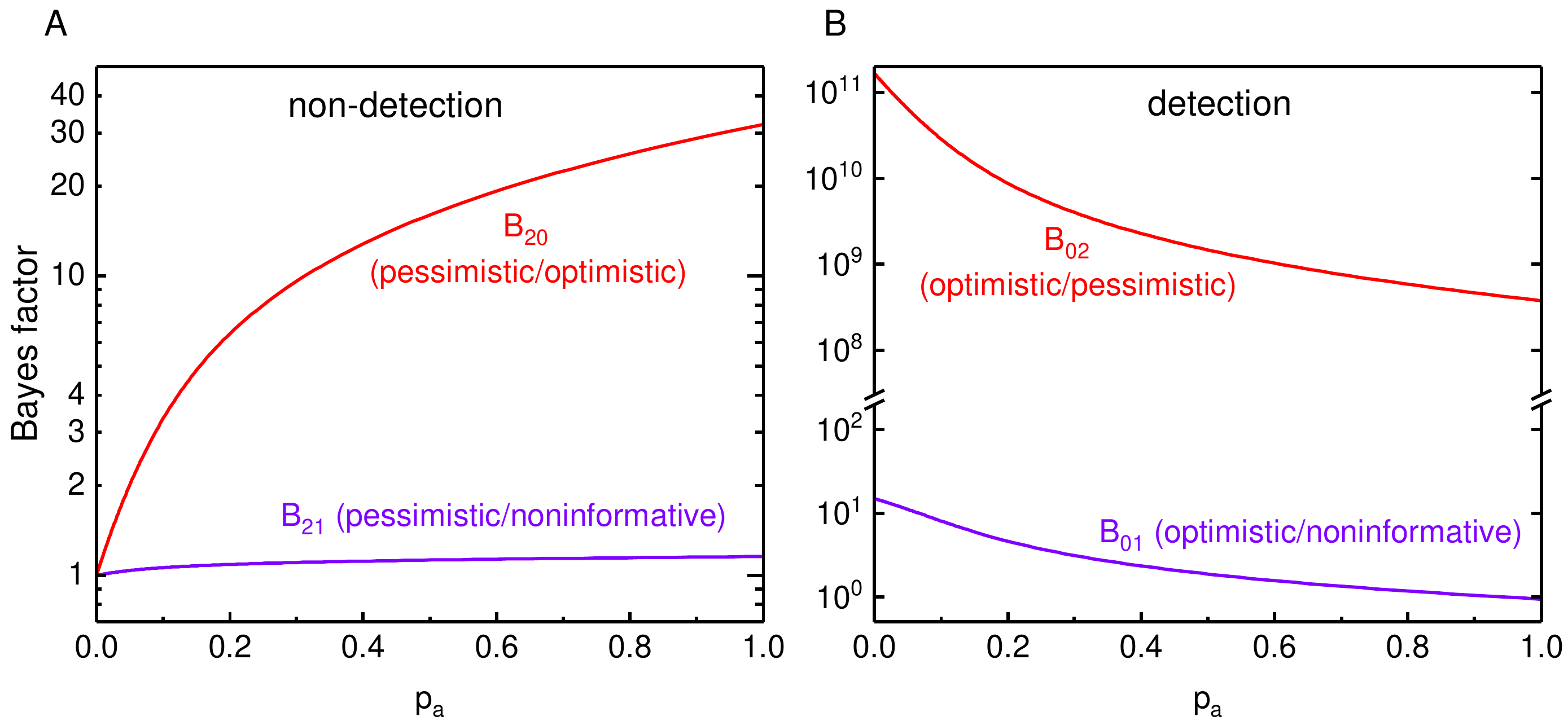}
		\caption{Bayes factor as a function of $p_a$, in a survey searching for atmospheric biosignatures within a distance $R=100$ ly. From left to right: (A) Bayes factor from the comparison of the pessimistic vs.\ optimistic model (red) and pessimistic vs.\ non-informative model (blue), when no biosignature detection is made. (B) Bayes factor from the comparison of the optimistic vs.\ pessimistic model (red) and optimistic vs.\ non-informative model (blue), when exactly one biosignature detection is made.  
		}\label{fig2}
	\end{center}
\end{figure*}

Figure \ref{fig1}A compares the impact of observing or not observing biosignatures within $100$ ly. In the case of non-detection, the posterior PDF of $\bar{k}$ differs only marginally from the log-uniform prior (long dashed line) in the range $\bar{k}\lesssim \pi(R)^{-1}\approx 10^6$, even assuming a complete survey ($p_{\rm a}=1$).
	The resulting complementary cumulative distribution function (CCDF) of $\bar{k}$ (Fig.~\ref{fig1}D) is somewhat smaller than the corresponding prior CCDF, the main deviation being
	an upper cutoff for $\bar{k}\gtrsim 10^6$, about $100$ times smaller than $\bar{k}_\textrm{max}$.
	 This limited response to non-detection is explained naturally by the smallness of $\pi(R)$ at $R=100$ ly, and it becomes even weaker as $p_{\rm a}$ diminishes, until the prior and posterior probabilities coincide in the entire $\bar{k}$ interval for $p_{\rm a}\rightarrow 0$. Therefore, even in the hypothesis that future surveys will rule out the existence of detectable biosignatures within $100$ ly, the added informative value will nevertheless remain modest, affecting only weakly the initial assertion of a non-informative, log-uniform prior. This conclusion is robust against a lowering of $\bar{k}_\textrm{min}$ and/or $\bar{k}_\textrm{max}$ (\textit{SI Appendix}, Fig. S8). In particular, reducing $\bar{k}_\textrm{max}$ below $\pi(R)^{-1}\approx 10^6$ is equivalent to assuming that planets with biosignatures are rare enough that finding none within such small survey volume is hardly surprising.

By contrast, the discovery of biosignatures on even a single planet within the entire survey volume ($R= 100$ ly, $p_{\rm a}=1$) would bring a response markedly different from the prior: we find a posterior PDF strongly peaked around $\bar{k}=3\times 10^6$, and a probability exceeding $95$ \% that $\bar{k}> 10^5$. For the sake of comparison, this would imply that exoplanet biosignatures, if distributed homogeneously throughout the Galaxy, are far more common than pulsar stars. Even larger values of $\bar{k}$ would be inferred by detecting a biosignature in a sample with few targeted planets, as illustrated by the limiting case $p_{\rm a}=0$  in Figs.~\ref{fig1}A and \ref{fig1}D (dotted lines) (\textit{SI Appendix}, section I). 
We further note that although changing $\bar{k}_\textrm{min}$ does not modify this conclusion, a detection event assuming $\bar{k}_\textrm{max}$ smaller than $\pi(R)^{-1}\approx 10^6$ would bring a response totally independent of the sample fraction $p_{\rm a}$, hinting to a larger $\bar{k}_\textrm{max}$ (\textit{SI Appendix}, Fig. S8). 

To provide a more complete analysis of the non-informed case, we have considered also the log log uniform prior, which has been designed to reflect total ignorance 
about the number of conditions conducive to life \cite{Lacki2016}. Although the log log uniform PDF slightly favors large values of $\bar{k}$, the resulting posteriors are in semi-quantitative agreement with those resulting from the log-uniform prior of Figs.~\ref{fig1}A and \ref{fig1}D (\textit{SI Appendix}, section IV).

\subsection*{Informative priors}

A log-uniform PDF is probably the best prior reflecting the lack of information on $\bar{k}$ even at the order-of-magnitude level. However, it is also worthwhile to explore how more informative prior distributions are updated once new evidence is gathered. Two interesting limiting cases are those reflecting a pessimistic or optimistic stance on the question of extraterrestrial life. On one hand, it has been argued that abiogenesis may result from complex chains of chemical reactions that have a negligibly low probability of occurring. Furthermore, contingent events which are thought to have favored an enduring biosphere on Earth (like, for example, a moon stabilizing the rotation axis of the planet, plate tectonic, etc.) may be so improbable to further lower the population of biosignature-bearing exoplanets. This view would result in a more pessimistic attitude toward the prior, with small values of $\bar{k}$ being preferred with respect to large ones. We model this case by adopting the uniform in $\bar{k}^{-1}$ prior $p(\bar{k}\vert M_2)\propto \bar{k}^{-2}$ in the interval $\bar{k}_{\rm min}$ to $\bar{k}_{\rm max}$. Conversely, the astronomically large number of rocky planets in the Milky Way combined with the assumption that the Earth is not special in any way (often termed `principle of mediocrity'), may suggest the optimistic hypothesis that life is very common in the Galaxy and the universe, resulting in a prior which weighs large values of $\bar{k}$ more favorably. We capture this view by taking the uniform in $\bar{k}$ prior $p(\bar{k}\vert M_0)$.

Figures \ref{fig1}B and C show the posterior PDFs and CCDFs resulting from detection or non-detection starting from a pessimistic hypothesis about $\bar{k}$. While the response to non-detection practically coincides with the prior expectation (an unsurprising result, given that the prior favors small values of $\bar{k}$) the event of detecting a biosignature increases the cut-off on $\bar{k}$ from $\sim 10^{-3}$ before the detection to at least $\sim 10^6$ after a biosignature is observed within the entire volume sample ($R=100$ ly, $p_{\rm a}=1$). In the optimistic model of Fig.~\ref{fig1}C and F the prior strongly constraints the posteriors resulting from both the events of detection and non-detection. In particular, the smallness of $\pi(R)$ shifts the CCDF resulting from the non-detection by a factor of only $\sim 10^{-1}$ in $\bar{k}$ (Fig.~\ref{fig1}F), not justifying thus a substantial revision of the initial optimistic stance.

\begin{figure*}[t]
	\begin{center}
		\includegraphics[width=0.9\textwidth,clip=true]{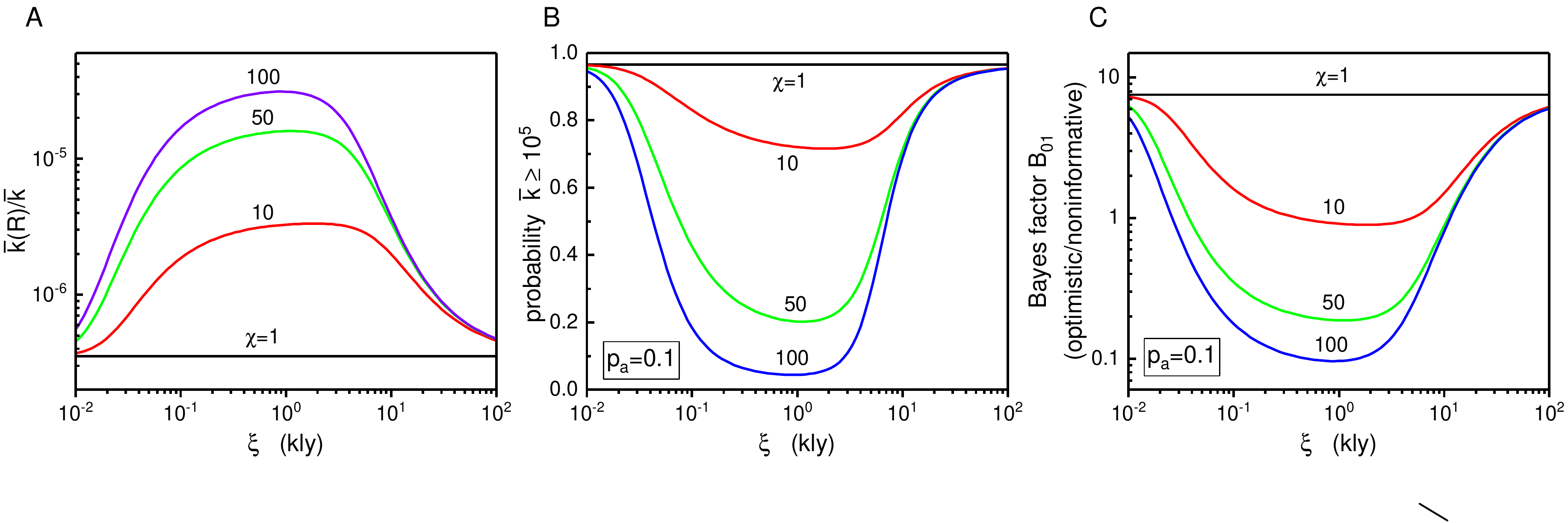}
		\caption{Effect of assuming correlation between biosignatures due to a panspermia mechanism, after a detection is made in a survey with $R=100$ ly and $p_a=0.1$. From left to right: (A) The fraction of life-harboring planets within a distance $R$ from Earth, $\bar{k}(R)$, over the total number $\bar{k}$ in the entire Galaxy, as a function of the correlation length $\xi$ and correlation strength $\chi$; (B) The probability that the total number of life-bearing planets in the Galaxy $\bar{k}$ is larger than a reference value $10^5$ in the entire Galaxy, as a function of the correlation length $\xi$; (C) Bayes factor from the comparison of the optimistic vs.\ non-informative model. 
		}\label{fig3}
	\end{center}
\end{figure*}

\subsection*{Model comparison} 
By adopting impartial judgement about the probability of $M_i$ being true ($i=0,\, 1,\, 2$),
we compute the Bayes factor $B_{ij}$ often used in model selection, giving the plausibility of model $M_i$ compared to $M_j$ in the face of the evidence (i.e.\ detection or non-detection):
\begin{equation}
\label{bayesM2}
B_{ij}(\mathcal{E}_k)=\frac{P(\mathcal{E}_k\vert M_i)}{P(\mathcal{E}_k\vert M_j)}.
\end{equation}
As a reference, $B_{ij}>10$ is usually considered as strong reason to prefer model $M_i$ over $M_j$. 

Model comparison through the Bayes factor (Figure \ref{fig2}) shows that if no detection is made, a pessimistic credence with regard to extraterrestrial life would strongly increase its likelihood with respect to an optimistic one, with a Bayes factor above 10, only if $p_{\rm a}$ is larger than 40\%.  The increase with respect to a neutral, non-informative stance would be, instead, basically insignificant for all $p_{\rm a}$ values. This teaches us that unless future surveys will search for biosignatures within a significant fraction of the  volume within $100$ ly (say, $p_{\rm a}>10$ \%) detecting none will not support convincingly either hypothesis. On the other hand, if a detection is made, the optimistic scenario would be hugely favored (Bayes factor larger than $10^8$) with respect to the pessimistic, and would be substantially preferable even with respect to a neutral position when $p_{\rm a}$ is close to 0. Somewhat counterintuitively, however, finding a single biosignature within a significant fraction of the volume $R=100$ ly (i.e.,\ $p_{\rm a}$ larger than $40-50$ \%) would not justify entirely the preference for an optimistic credence compared to the non-informative hypothesis. These results put on a quantitative and rigorous statistical basis the common intuitive idea that the discovery of even a single unambiguous biosignature would radically change our attitude towards the frequency of extraterrestrial life.

\subsection*{Impact of panspermia scenarios}

So far, we have assumed that any given planet has some probability of harboring life independently of whether or not other planets harbor life as well. However, in general this may not be the case. For example, according to the hypothetical panspermia scenario, life might be transferred among planets, within the same stellar system, in stellar clusters, or over interstellar distances \cite{Melosh2003,Adams2005,Ginsburg2018,Lin2015}. If conditions favor the flourishing of a biosphere within a relatively short time-scale after the transfer, this would result in an enhanced probability that a planet is inhabited if a nearby planet is inhabited as well \cite{Lingam2016}. In this way, if panspermia can occur, the probability that two planets produce simultaneously biosignatures will depend on their relative distance and on a typical length scale, that we denote $\xi$, defined by the capability of life of surviving transfer and establishing a biosphere.  

We took this possibility into account by modeling the statistical correlation of biosignatures and rewriting $\bar{k}(R)$ as follows:\begin{equation}
\label{pan1}
\bar{k}(R)=p\int\!d\mathbf{r}\rho(\mathbf{r})g(\mathbf{r},\mathbf{r}_\textrm{E})\theta(R-\vert\mathbf{r}-\mathbf{r}_\textrm{E}\vert),
\end{equation} 
where the pair distribution function $g(\mathbf{r},\mathbf{r}')$ gives the relative
probability of biosignatures being present on $\mathbf{r}$ if biosignatures are present also in $\mathbf{r}'$.

In principle, different models of correlation could be linked to specific panspermia mechanisms, and various scenarios might even be distinguished observationally from an independent abiogenesis \cite{Lingam2016}. This could be an interesting subject for future studies. However, here we are only interested in how the presence of generic correlations would impact the statistical significance of biosignatures detection.  We adopt a simple model for the pair distribution function, by assuming that it depends on the relative distance $\vert\mathbf{r}-\mathbf{r}'\vert$ in such a way that:
\begin{equation}
\label{rdf2}
g(\vert\mathbf{r}-\mathbf{r}'\vert)=(\chi-1)e^{-\vert\mathbf{r}-\mathbf{r}'\vert/\xi}+1,
\end{equation}
where $\chi\geq 1$ describes the intensity of the panspermia process and $\xi$ its spatial extension. The uncorrelated case $g(\vert\mathbf{r}-\mathbf{r}'\vert)=1$ 
(no panspermia) is obtained by setting $\chi=1$, while
$\chi\gg 1$ yields a strong probability of finding two life-harboring planets within a relative distance $\lesssim \xi$ from each other. 
Using $p=p_{\rm a}p_{\rm d}p_{\rm l}$, the average number of exoplanet biosignatures in the entire galaxy is given by $\bar{k}=p_{\rm d}p_{\rm l}\int\! d\mathbf{r}\rho(\mathbf{r})g(\vert\mathbf{r}-\mathbf{r}_\textrm{E}\vert)$, so that \eqref{pan1} reduces to
\begin{equation}
\label{pan2}\bar{k}(R)=\bar{k}p_{\rm a}\frac{\int\!d\mathbf{r}\rho(\mathbf{r})g(\vert\mathbf{r}-\mathbf{r}_\textrm{E}\vert)\theta(R-\vert\mathbf{r}-\mathbf{r}_\textrm{E}\vert)}
{\int\!d\mathbf{r}\rho(\mathbf{r})g(\vert\mathbf{r}-\mathbf{r}_\textrm{E}\vert)}.
\end{equation}
The parameter $\chi$ is not unbounded, as within any radius $R$ there cannot be more planets with biosignatures than the total number of planets, 
$N(R)=N\pi(R)$, contained within $R$. In other words: $\bar{k}(R)\leq N(R)$ for any $R$. For $R< 1$ kly, this condition is automatically satisfied by 
imposing $\chi\leq N/\bar{k}_\textrm{max}$, where $\bar{k}_\textrm{max}/N$ 
is the maximum fraction of exoplanets harboring biosignatures. 

As shown in Fig.~\ref{fig3}A, the average number of biosignatures within $R$, $\bar{k}(R)$, gets enhanced by the panspermia mechanism with respect to the uncorrelated case, with $\bar{k}(R)/\bar{k}$ showing a broad maximum around $\xi=10^3$ ly (\textit{SI Appendix}, section IIIB). For much larger values of $\xi$, panspermia would distribute life homogeneously through the entire galaxy and $\bar{k}(R)/\bar{k}$ would approach unity.

Figure \ref{fig3}B shows that if possible correlations in the biosignatures are taken into account, the probability that the number of life-bearing planets in the galaxy is larger than a given value (taken as $10^5$ for the sake of illustration) decreases substantially even if life is detected in an incomplete sample (with $p_a=0.1$) in the volume $R=100$ ly. This shows as a decrease in the Bayes factor (Figure \ref{fig3}C) of the optimistic scenario with respect to the non-informative one. Depending on $\xi$ and $\chi$, there may be no gain in knowledge when life is detected elsewhere and for $\xi\approx 10^3$ ly a complete correlation ($\chi=100$) would even strongly favor the non-informative hypothesis over the optimistic one. This conclusion suggests that the viability of the panspermia scenario should be assessed independently (for example through experimental studies of the survivability of organisms in deep space), in order not to weaken the significance of the possible discovery of life beyond Earth.

\acknowledgements
A.B.\ acknowledges support by the Italian Space Agency (ASI, DC-VUM-2017-034, grant number 2019-3 U.O Life in Space) and by grant number FQXi-MGA-1801 and FQXi-MGB-1924 from the Foundational Questions Institute and Fetzer Franklin Fund, a donor advised fund of Silicon Valley Community Foundation.

\bibliography{biblio}

\cleardoublepage
\onecolumngrid
\appendix
%%%%%%%%%%%%%%%%%%%%%%%%%%%%%%%%%%%%
%%%APPENDIX
%%%%%%%%%%%%%%%%%%%%%%%%%%%%%%%%%%%%
%\section*{Supplementary material}
%\label{AppA}
%\input{Supplementary_material}


\section*{Supplementary material}

# Supplementary information for
# Quantifying the information impact of future searches for exoplanetary biosignatures

Amedeo Balbi[1, *] and Claudio Grimaldi[2, 3, †]

[1]*Dipartimento di Fisica, Università di Roma "Tor Vergata",
Via della Ricerca Scientifica, 00133 Roma, Italy*
[2]*Laboratory of Physics of Complex Matter, Ecole Polytechnique
Fédérale de Lausanne, Station 3, CP-1015 Lausanne, Switzerland*
[3]*Centro Studi e Ricerche Enrico Fermi, Piazza del Viminale, 1, 00184 Roma, Italy*

## I. POSTERIOR PDFS

We denote $\mathcal{E}_k$ the event of detecting exactly $k = 0, 1, 2, \ldots$ biosignatures during a survey parametrized by $p_a \pi(R)$, where $0 \leq p_a \leq 1$ takes into account astrophysical factors and observational limitation and $\pi(R)$ is the probability of a habitable planet being within a distance $R$ from Earth. The likelihood of $\mathcal{E}_k$ being true under the condition that in the entire galaxy there are in average $\bar{k}$ exoplanets producing biosignatures is:

$$P(\mathcal{E}_k|\bar{k}) = \frac{[\bar{k} p_a \pi(R)]^k}{k!} e^{-\bar{k} p_a \pi(R)}, \tag{S1}$$

so that using Bayes' theorem we infer the posterior PDF of $\bar{k}$ given $\mathcal{E}_k$:

$$p(\bar{k}|\mathcal{E}_k, M) = \frac{P(\mathcal{E}_k|\bar{k}) p(\bar{k}|M)}{P(\mathcal{E}_k|M)}, \tag{S2}$$

where $p(\bar{k}|M)$ is the prior PDF of $\bar{k}$ whose specific functional form is chosen according to the model $M$ and

$$P(\mathcal{E}_k|M) = \int d\bar{k} P(\mathcal{E}_k|\bar{k}) p(\bar{k}|M) \tag{S3}$$

is the likelihood of $\mathcal{E}_k$ given the model $M$.

We adopt a power-law form for the prior PDFs with three values of the exponent, $p(\bar{k}|M_i) \propto \bar{k}^{-i}$ (with $i = 0, 1, 2$), defined in the interval $\bar{k}_{\min}$ to $\bar{k}_{\max}$. Their normalized forms are:

$$\begin{aligned} p(\bar{k}|M_0) &= \frac{1}{\bar{k}_{\max} - \bar{k}_{\min}}, && \text{optimistic model } M_0 \\ p(\bar{k}|M_1) &= \frac{\bar{k}^{-1}}{\ln(\bar{k}_{\max}/\bar{k}_{\min})}, && \text{noninformative model } M_1 \\ p(\bar{k}|M_2) &= \frac{\bar{k}_{\max} \bar{k}_{\min}}{\bar{k}_{\max} - \bar{k}_{\min}} \bar{k}^{-2}, && \text{pessimistic model } M_1. \end{aligned} \tag{S4}$$

In the article, we consider two events: non-detection ($\mathcal{E}_0$) and the detection of exactly one biosignature ($\mathcal{E}_1$). From (S1) the corresponding likelihood functions are:

$$P(\mathcal{E}_0|\bar{k}) = e^{-\bar{k} p_a \pi(R)}, \qquad P(\mathcal{E}_1|\bar{k}) = \bar{k} p_a \pi(R) e^{-\bar{k} p_a \pi(R)}, \tag{S5}$$

which using (S2) and (S3) yield the following posterior PDFs:

---

* amedeo.balbi@roma2.infn.it          † claudio.grimaldi@epfl.ch

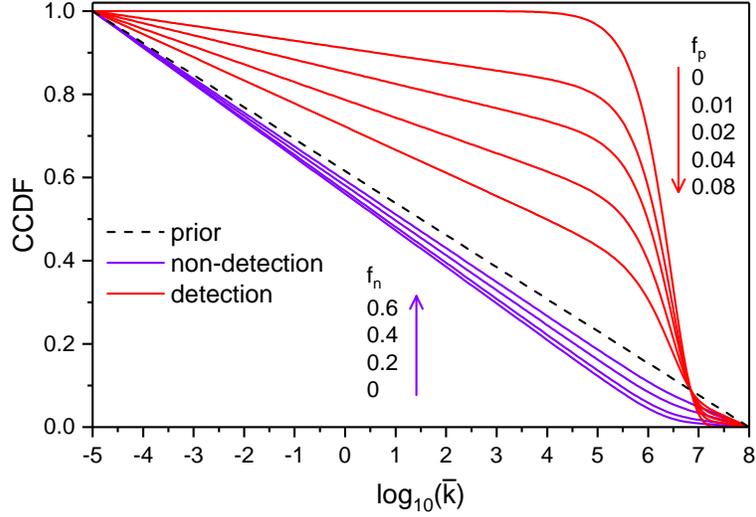

FIG. S1. CCDFs resulting from a search of biosignatures in the atmosphere of exoplanets within 100 ly from Earth, assuming a non-informative log-uniform prior (dashed line) and $p_a = 1$. False negatives and false positives are parametrized by the probability factors $f_n$ and $f_p$, respectively.

optimistic model $M_0$

$$p(\bar{k}|\mathcal{E}_0, M_0) = \frac{p_a \pi(R) e^{-p_a \pi(R) \bar{k}}}{e^{-p_a \pi(R) \bar{k}_{\min}} - e^{-p_a \pi(R) \bar{k}_{\max}}} \tag{S6}$$

$$p(\bar{k}|\mathcal{E}_1, M_0) = \frac{\bar{k}[p_a \pi(R)]^2 e^{-p_a \pi(R) \bar{k}}}{[1 + p_a \pi(R) \bar{k}_{\min}] e^{-p_a \pi(R) \bar{k}_{\min}} - [1 + p_a \pi(R) \bar{k}_{\max}] e^{-p_a \pi(R) \bar{k}_{\max}}} \tag{S7}$$

non-informative model $M_1$

$$p(\bar{k}|\mathcal{E}_0, M_1) = \frac{\bar{k}^{-1} e^{-p_a \pi(R) \bar{k}}}{E_1[p_a \pi(R) \bar{k}_{\min}] - E_1[p_a \pi(R) \bar{k}_{\max}]} \tag{S8}$$

$$p(\bar{k}|\mathcal{E}_1, M_1) = \frac{p_a \pi(R) e^{-p_a \pi(R) \bar{k}}}{e^{-p_a \pi(R) \bar{k}_{\min}} - e^{-p_a \pi(R) \bar{k}_{\max}}} \tag{S9}$$

pessimistic model $M_2$

$$p(\bar{k}|\mathcal{E}_0, M_2) = \frac{\bar{k}^{-2} e^{-p_a \pi(R) \bar{k}}}{\bar{k}_{\min}^{-1} e^{-p_a \pi(R) \bar{k}_{\min}} - \bar{k}_{\max}^{-1} e^{-p_a \pi(R) \bar{k}_{\max}} - p_a \pi(R) \{E_1[p_a \pi(R) \bar{k}_{\min}] - E_1[p_a \pi(R) \bar{k}_{\max}]\}} \tag{S10}$$

$$p(\bar{k}|\mathcal{E}_1, M_2) = \frac{\bar{k}^{-1} e^{-p_a \pi(R) \bar{k}}}{E_1[p_a \pi(R) \bar{k}_{\min}] - E_1[p_a \pi(R) \bar{k}_{\max}]}, \tag{S11}$$

where $E_1(x) = \int_x^\infty dx \, \exp(-x)/x$ is the exponential integral function. Note that the $p_a \to 0$ limit of $p(\bar{k}|\mathcal{E}_0, M_i)$ yields the corresponding prior PDFs of (S4), while the $p_a \to 0$ limit of $p(\bar{k}|\mathcal{E}_1, M_i)$ reduces to

$$\lim_{p_a \to 0} p(\bar{k}|\mathcal{E}_1, M_0) = \frac{2\bar{k}}{\bar{k}_{\max}^2 - \bar{k}_{\min}^2}, \quad \lim_{p_a \to 0} p(\bar{k}|\mathcal{E}_1, M_1) = \frac{1}{\bar{k}_{\max} - \bar{k}_{\min}}, \quad \lim_{p_a \to 0} p(\bar{k}|\mathcal{E}_1, M_2) = \frac{\bar{k}^{-1}}{\ln(\bar{k}_{\max}/\bar{k}_{\min})}. \tag{S12}$$

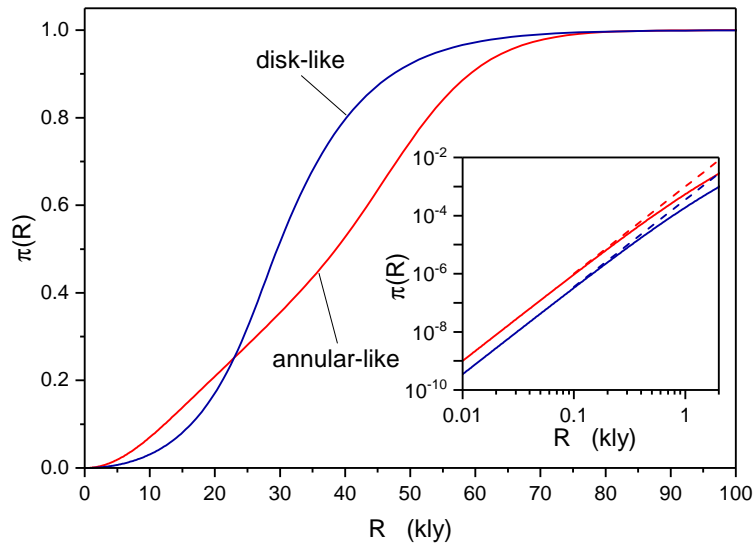

FIG. S2. Probability $\pi(R)$ of an habitable exoplanet being within a distance $R$ from Earth for a disk-like and an annular-like spatial distribution of exoplanets. The inset shows that within the galactic neighbourhood ($R \lesssim 1$ kly) $\pi(R)$ scales as $(R/a)^3$ (dashed lines) where $a = 14.2$ kly and $a = 10$ kly for disk-like and annular-like distributions, respectively.

## II. FALSE POSITIVE AND FALSE NEGATIVE EVENTS

The possibility of both false negatives (biosignatures that are present but go undetected) and false positives (gases of abiotic origin that are mistakenly interpreted as products of life) makes the search for biosignatures prone to ambiguous results. To account for such ambiguity in a simple way, we introduce the probability $f_n$ of having false negative non-detections and the probability $f_p$ of false positive detections. Assuming a systematic failure in recognizing both true detections and true non-detections, we follow Ref. [S1] and express the likelihood functions as

$$P(\mathcal{E}_0, f_n|\bar{k}) = f_n + (1 - f_n)e^{-\bar{k}p_a\pi(R)}, \qquad P(\mathcal{E}_1, f_p|\bar{k}) = f_p + (1 - f_p)\bar{k}p_a\pi(R)e^{-\bar{k}p_a\pi(R)}. \tag{S13}$$

In this way, the updated PDFs are just a linear combination of the prior and the posterior PDF obtained by ignoring the systematic failure. In Fig. S1 we show the CCDFs resulting from the non-detection and the detection of biosignatures within $R = 100$ ly from the Earth assuming the log-uniform prior (non-informed model $M_1$, dashed line) and $p_a = 1$. The results for $f_n = f_p = 0$ correspond to the absence of false positives and negatives and coincide with those plotted in Fig. 1d of the main text. As $f_n$ and $f_p$ increase, the corresponding posterior CCDFs tend gradually towards the prior probability. Interestingly, the effect of $f_p$ is much stronger than that of $f_n$. This is understood by the marginal added information from the non-detection of biosignatures within the small sample volume covered by the survey ($\pi(100\text{ly}) \simeq 3.2 \times 10^{-7}$).

## III. MODELS FOR THE EXOPLANET NUMBER DENSITY AND CORRESPONDING $\pi(R)$

The probability of a habitable exoplanet being within a distance $R$ from Earth is:

$$\pi(R) = N^{-1} \int d\mathbf{r}\, \rho(\mathbf{r}) \theta(R_\mathsf{L} - |\mathbf{r} - \mathbf{r}_\mathrm{E}|), \tag{S14}$$

where $\rho(\mathbf{r})$ is the number density of habitable exoplanets in the Galaxy, $N$ is their number, $\mathbf{r}_\mathrm{E}$ is the position vector of the Earth relative to the galactic center, and $\theta(x)$ is the Heaviside step function.

We consider a general axisymmetric model for $\rho(\mathbf{r})$:

$$\rho(\mathbf{r}) = N \frac{(r/r_s)^\beta e^{-r/r_s} e^{-|z|/z_s}}{4\pi r_s^2 z_s \Gamma(\beta + 2)}, \tag{S15}$$

where $\beta \geq 0$, $r$ is the radial distance from the galactic center, $z$ is the height from the galactic plane, and $\Gamma(x)$ is the Gamma function. Depending on the value of $\beta$ and $r_s$, the radial dependence of $\rho(\mathbf{r})$ can be adjusted so as

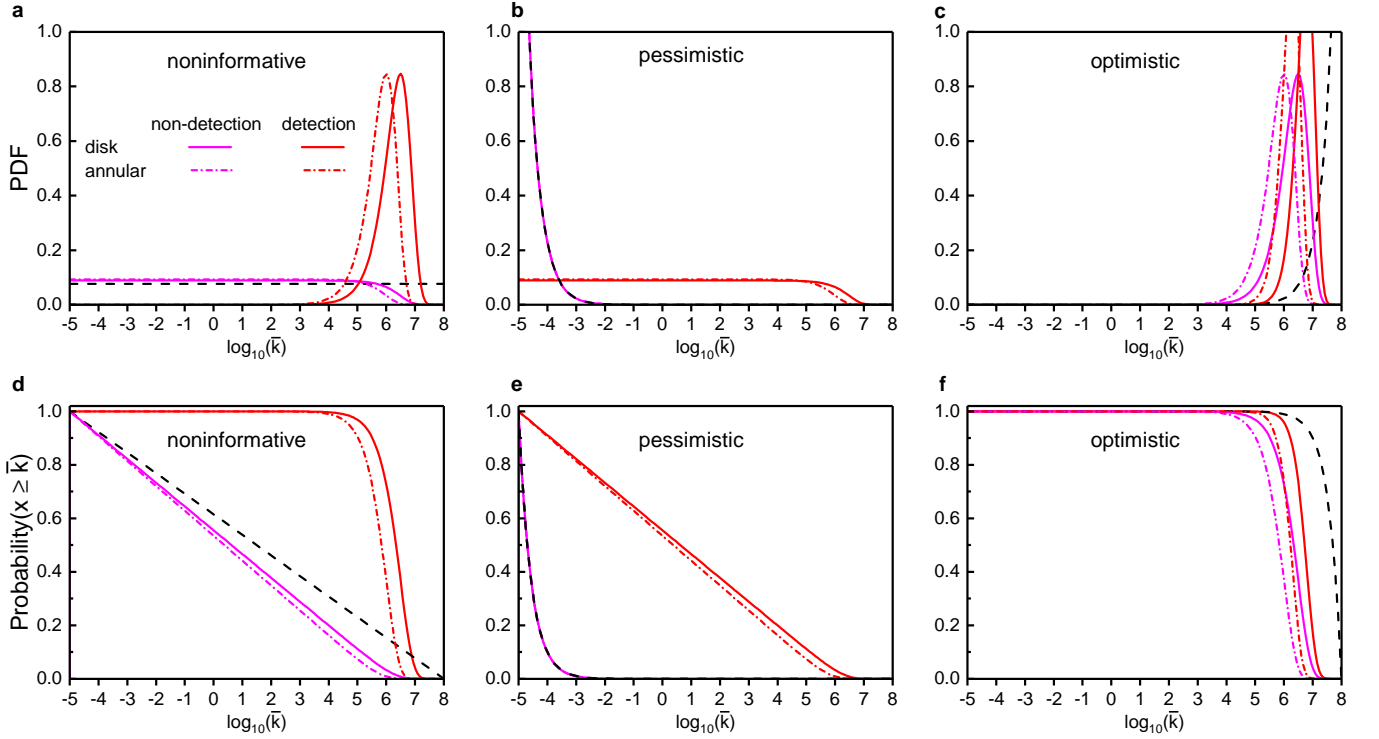

FIG. S3. PDFs (upper row) and CCDFs (bottom row) resulting from a survey within $R = 100$ ly calculated for disk-like (solid lines) and annular-like (dot-dashed lines) galactic habitable zones (GHZs). The probability $\pi(R = 100\,\text{ly})$ has been calculated using (S17) and (S18) resulting in $\pi(100\,\text{ly}) = 3.2 \times 10^{-7}$ and $10^{-6}$ for the disk-like and annular-like GHZ, respectively. Noninformative, pessimistic and optimistic priors are plotted by black dashed lines in the interval $\bar{k}_{\min} = 10^{-5}$ to $\bar{k}_{\max} = 10^8$. Only the posteriors calculated with $p_a = 1$ are shown because for $p_a \to 0$ the form of $\pi(R)$ is irrelevant.

to reproduce different galactic distributions of habitable exoplanets. Here, we assume that $\rho(\mathbf{r})$ follows the density profile of the galactic habitable zone (GHZ) which takes into account factors that are thought to be important for the development of life, such as the star metallicity gradient and the rate of major sterilizing events (e.g., supernovae). In the article we have considered a GHZ extending over the entire thin disk of the Galaxy (disk-like GHZ) by taking $\beta = 0$, $r_s = 8.15$ kly, and $z_s = 0.52$ kly [S2]. An alternate model of $\rho(\mathbf{r})$ is the annular GHZ of ref. [S3] which can be reproduced by taking $\beta = 7$, $r_s = 3.26$ kly, and $z_s = 0.52$ kly (annular-like GHZ). These parameters give a probability of 68 % of finding stars with the highest potential of harboring complex life within $\approx 15$ kly and $\approx 34$ kly from the galactic center, which is a good match with the time averaged GHZ of ref. [S3].

Fig. S2 shows the probability $\pi(R)$ as a function of $R$ calculated for these two models: the probability of finding a habitable planet within $\sim 25$ kly from Earth is larger for the annular-like GHZ while for $R \gtrsim 25$ kly the disk-like model yields a greater $\pi(R)$.

For $R$ much smaller than the typical length scale over which $\rho(\mathbf{r})$ varies, (S14) can be Taylor expanded for small $R$:

$$\pi(R) \simeq \frac{\rho(\mathbf{r}_\mathrm{E})}{N} \int d\mathbf{r}\, \theta(R - |\mathbf{r} - \mathbf{r}_\mathrm{E}|) = \frac{4\pi}{3}\frac{\rho(\mathbf{r}_\mathrm{E})}{N} R^3. \tag{S16}$$

We take the Earth to be located approximately on the galactic plane, $\mathbf{r}_\mathrm{E} = (r_\mathrm{E}, 0)$, with its radial distance from the center of the Milky Way being $r_\mathrm{E} = 27$ kly. We obtain from Eqs. S16 and S15:

$$\pi(R) \simeq \frac{(r_\mathrm{E}/r_s)^\beta e^{-r_\mathrm{E}/r_s}}{3 r_s^2 z_s \Gamma(\beta+2)} R^3 = \left(\frac{R}{a}\right)^3, \tag{S17}$$

where

$$a = \begin{cases} 14.2\ \text{kly}, & \text{disk-like GHZ}, \\ 10\ \text{kly}, & \text{annular-like GHZ}. \end{cases} \tag{S18}$$

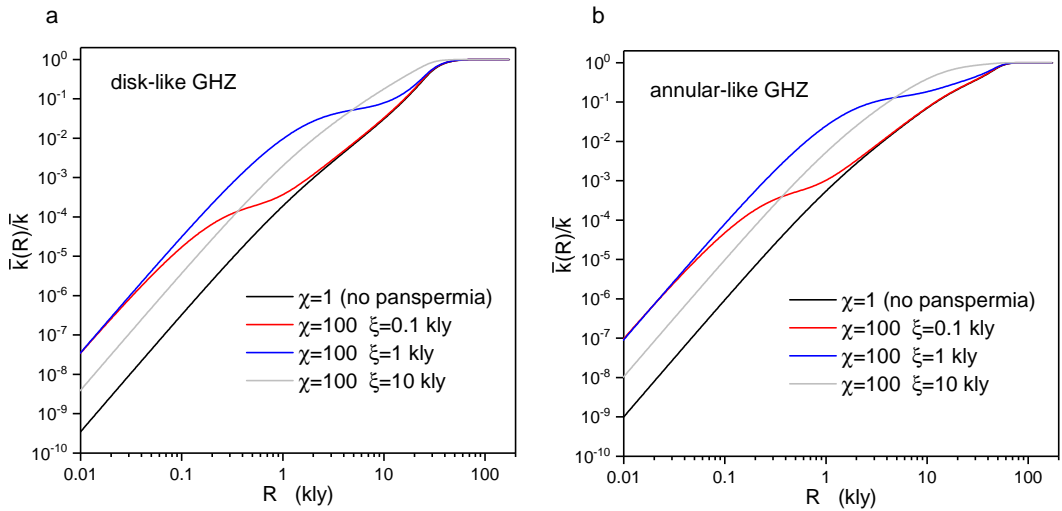

FIG. S4. $\bar{k}(R)/\bar{k}$ is the fraction of biosignature-producing planets within a distance $R$ from Earth over the total number $\bar{k}$ in the entire Galaxy ($p_a = 1$ is assumed). Panels a and b show the results obtained from the disk-like and the annular-like GHZs, respectively. The curves for $\chi = 1$ correspond to the case without panspermia and coincides with $\pi(R)$, the probability of an habitable exoplanet being within $R$. As the correlation length $\xi$ increases beyond 10 kly, $\bar{k}(R)/\bar{k}$ asymptotically reduces to the homogeneous case $\chi = 1$.

### A. Comparison of posteriors resulting from disk-like and annular-like GHZs

The effects of the habitable exoplanet distribution on the posterior probabilities for the noninformative, pessimistic, and optimistic model are shown in Fig. S3 for $R = 100$ ly. The posterior PDFs resulting from the annular-like GHZ (dot-dashed lines) are slightly shifted to lower values of $\bar{k}$ with respect to the PDFs of the disk-like model (solid lines). This is due to the larger value of $\pi(100\,\mathrm{ly})$ ($\simeq 10^{-6}$) for the annular GHZ compared to that resulting from a disk-like GHZ ($\simeq 3.2 \times 10^{-7}$).

### B. Effects of GHZ on panspermia correlations

The form of the GHZ affects also the exoplanet distribution resulting from the panspermia scenario. The log-log plots of Fig. S4 show the number of biosignature-producing exoplanet within a distance $R$ from Earth, $\bar{k}(R)$, over the total number $\bar{k}$ in the entire Galaxy (with $p_a = 1$) for different values of the correlation length $\xi$ and for panspermia intensity $\chi = 1$, corresponding to the scenario without panspermia mechanism, and $\chi = 100$, which is the maximum value allowed. Figures S4a and S4b show the results obtained by assuming the disk-like model and the annular-like model of the GHZ, respectively. Since the density $\rho(\mathbf{r})$ in the neighborhood of the solar system of the annular GHZ is larger than that of the disk-like GHZ, the value in that region of $\bar{k}(R)/\bar{k}$ corresponding to the annular model is larger than that of the disk-like model even in the presence of the panspermia mechanism.

The difference between the disk-like and the annular-like GHZs is however not so important to imply significant quantitative changes in the panspermia effects on the posterior probabilities. This is shown in Fig. S5 where $\bar{k}(R)/\bar{k}$, the probability of $\bar{k} > 10^5$ after a detection is made (event $\mathcal{E}_1$, non-informative model), and the Bayes factor from the comparison of the optimistic and the non-informative model, are plotted from left to right as a function of the correlation length $\xi$ for the disk-like GHZ (upper row, the same as in Fig. 3 of the same text) and the annular-like GHZ (lower row). Among the quantitative differences induced by the annular GHZ, it is worth to mention the 0% probability of $\bar{k}$ being larger than $10^5$ (middle column) for $\chi > 50$ and $\xi \approx 1$ kly,

## IV. COMPARISON BETWEEN LOG-UNIFORM, LOG LOG-UNIFORM, AND JEFFREY PRIORS

The log-uniform prior $p(\log \bar{k}) = \mathrm{const.}$, which corresponds to the non-informed model $p(\bar{k}|M_1) \propto 1/\bar{k}$ used in the article, gives equal weight to all orders of magnitude of $\bar{k}$, reflecting thus the total lack of knowledge about even the scale of $\bar{k}$. Recalling that we have defined in the article that the mean number of planets producing biosignatures

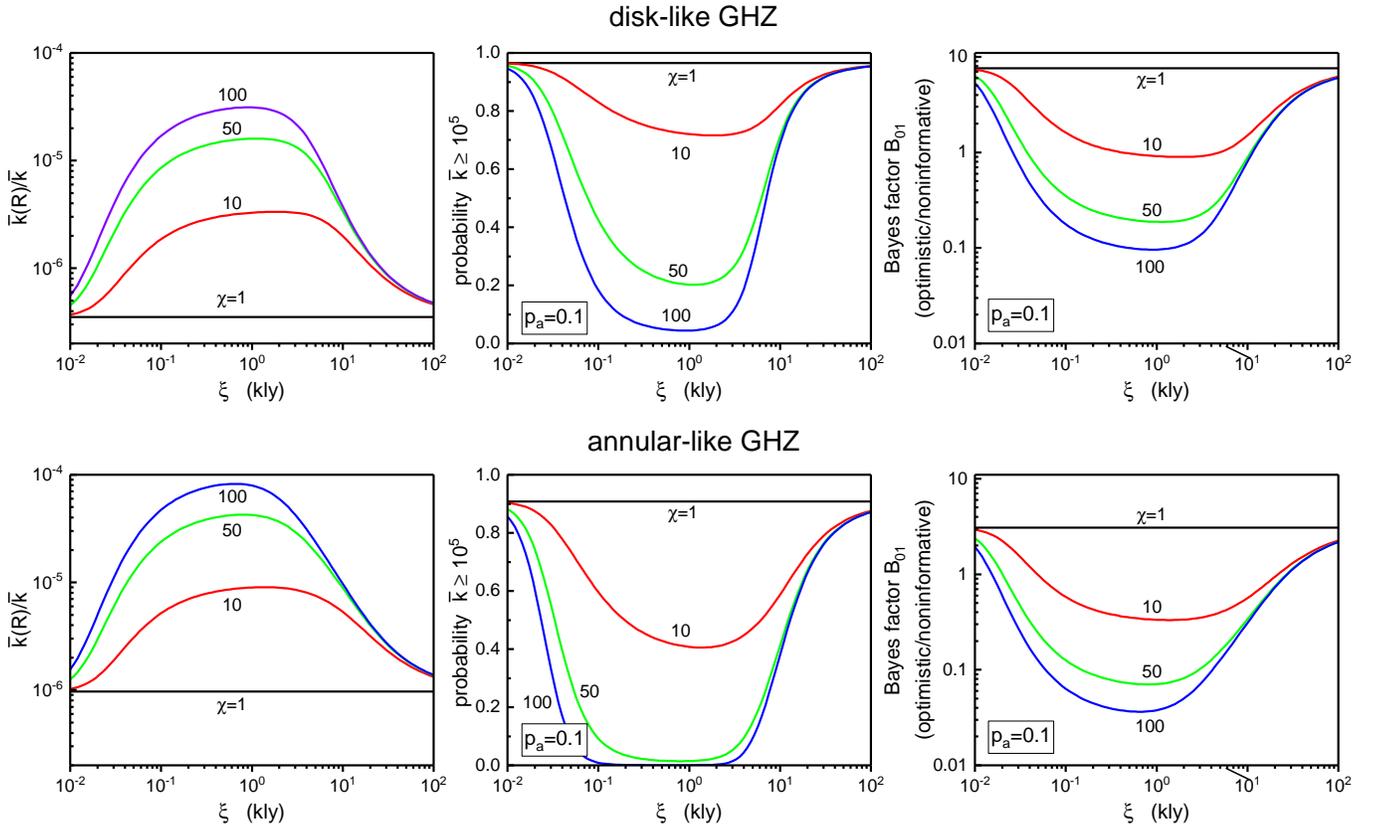

FIG. S5. Effect of assuming correlation between biosignatures due to a panspermia mechanism, after a detection is made in a survey with $R = 100$ ly and $p_a = 0.1$. The upper and bottom rows show the results obtained from assuming the disk-like GHZ (as in Fig. 3 of the main text) and the annular-like GHZ, respectively. From left to right: the fraction of life-harboring planets within a distance $R$ from Earth, $\bar{k}(R)$, over the total number $\bar{k}$ in the entire Galaxy, as a function of the correlation length $\xi$ and correlation strength $\chi$; the probability that the total number of life-bearing planets in the Galaxy $\bar{k}$ is larger than a reference value $10^5$ in the entire Galaxy, as a function of the correlation length $\xi$; Bayes factor from the comparison of the optimistic vs. non-informative model.

is $\bar{k} = p_d p_l N$, where $N$ is the number of habitable exoplanets, the total lack of knowledge has to be ascribed to the product probability $q = p_d p_l$, which is the probability that a habitable exoplanet produces biosignatures.

Recently, Lacki introduced an alternative non-informed prior specially designed to reflect our ignorance about the number of conditions that must be fulfilled for extraterrestrial life to emerge with probability $q$ on a habitable planet [S4]. Supposing that there are $n$ conditions independent of each other and that $n$ is uncertain at the order of magnitude level, then the prior PDF for $n$ should be constant in $\log n$, which translates into a prior for $q$ that is constant in $\log |\log q|$. To avoid an unphysical divergence at $q = 1$, Ref. [S4] introduced a prior that is flat in $\ln(1 - \ln q)$ which corresponds to a PDF of $q$ having the functional form [here we are using the natural logarithm to conform with the notation of Ref. [S4]]:

$$p_{\text{loglog}}(q) \propto \frac{1}{q(1 - \ln q)}, \tag{S19}$$

which, as shown in Fig. S6, puts more weight over larger values of $q$ than the log-uniform PDF $\propto 1/q$ (dashed horizontal line). This behavior is confirmed by an explicit calculation of the log log uniform prior obtained by taking $q$ equal to the product of $n \geq 1$ independent probabilities uniformly distributed in the interval 0 to 1, whose PDF is given by:

$$\rho_n(q) = \int_0^1 dp_1 \int_0^1 dp_2 \cdots \int_0^1 dq_n \, \delta\!\left(q - \prod_{i=1}^n p_i\right), \tag{S20}$$

where $\delta(x)$ is the Dirac delta function. (S20) can be calculated iteratively, yielding $\rho_n(q) = [\ln(1/q)]^{(n-1)}/(n-1)!$.

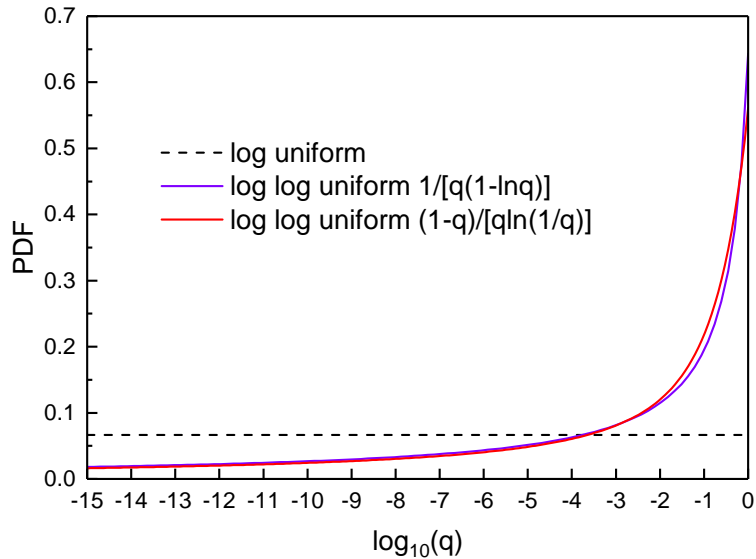

FIG. S6. Comparison of noninformed priors for $q$ defined in the interval $10^{-15}$ to 1 and properly renormalized. Compared to the log-uniform prior (dashed line), the weights of the log log uniform priors (S19) (red solid line) and (S21) (violet solid line) are somewhat shifted towards the large values of $q$.

Assuming a log-uniform distribution of $n$ we find

$$p_{\text{loglog}}(q) \propto \sum_{n=1}^{\infty} \frac{1}{n} \frac{[\ln(1/q)]^{n-1}}{(n-1)!} = \frac{1-q}{q\ln(1/q)}, \tag{S21}$$

which nicely agrees with the prior of (S19) shown in Fig. S6 (in passing we note that if $n$ is distributed uniformly over $[1, +\infty)$ we recover the log-uniform distribution $\propto 1/q$).

The prior of $\bar{k}$ corresponding to the log log uniform PDF is obtained from $p(\bar{k}) = N^{-1} p_{\text{loglog}}(\bar{k}/N)$, where $N = 10^{10}$ is the number of habitable exoplanets. Figure S7 shows the calculated posterior PDFs (Fig. S7b) and CCDFs (Fig. S7e) of $\bar{k}$ resulting from either the non-detection or the detection of exactly one biosignature in planets within 100 ly from the Earth. The overall behavior of the posteriors is qualitatively similar to that obtained in the case of a log-uniform prior (Figs- S7a and S7d), with only the weight of the posterior PDFs being slightly shifted towards large values of $\bar{k}$, as expected from the similar weight shift observed in the log log uniform prior (Fig. S6).

As another example of non-informative prior, we have calculated the posteriors resulting from the detection and the non-detection of biosignatures obtained by adopting the Jeffreys prior, which has the property of being invariant under a re-representation of the parameter $\bar{k}$. In the case treated in the article of a Poisson distribution of the number of planets producing biosignatures, the relevant Jeffreys prior of $\bar{k}$ is proportional to $1/\sqrt{\bar{k}}$. It is therefore a model distribution falling in between the non-informative log-uniform distribution, $p(\bar{k}|M_1) \propto 1/\bar{k}$, and the optimistic distribution $p(\bar{k}|M_2) = $ constant. Figures. S7c and S7f show indeed that the posterior PDFs have their weights significantly shifted towards large values of $\bar{k}$, although not so strongly shifted as in the optimistic model (Fig. 1C of the article).

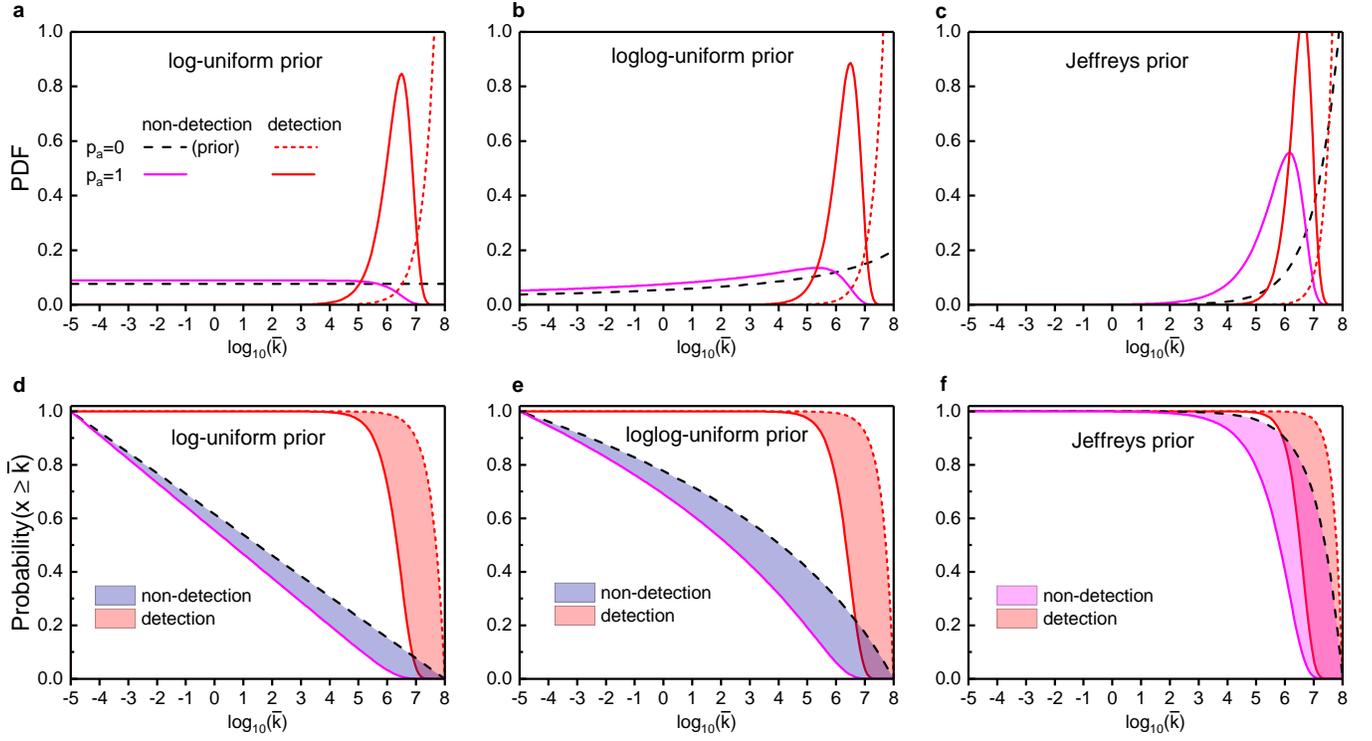

FIG. S7. PDFs (upper row) and CCDFs (bottom row) resulting from a survey within a distance $R = 100$ ly from Earth. Shown in each row, from left to right, are the posterior PDF (upper row) and the CCDF (bottom row) starting from respectively the log-uniform, the log log uniform, and the Jeffreys prior ($\propto 1/\sqrt{\bar{k}}$). Solid lines refer to posteriors calculated for $p_a = 1$, while the red short dashed lines refer to the case $p_a = 0$. In the case of non-detection, the posteriors obtained in the limit $p_a = 0$ coincide with the prior.

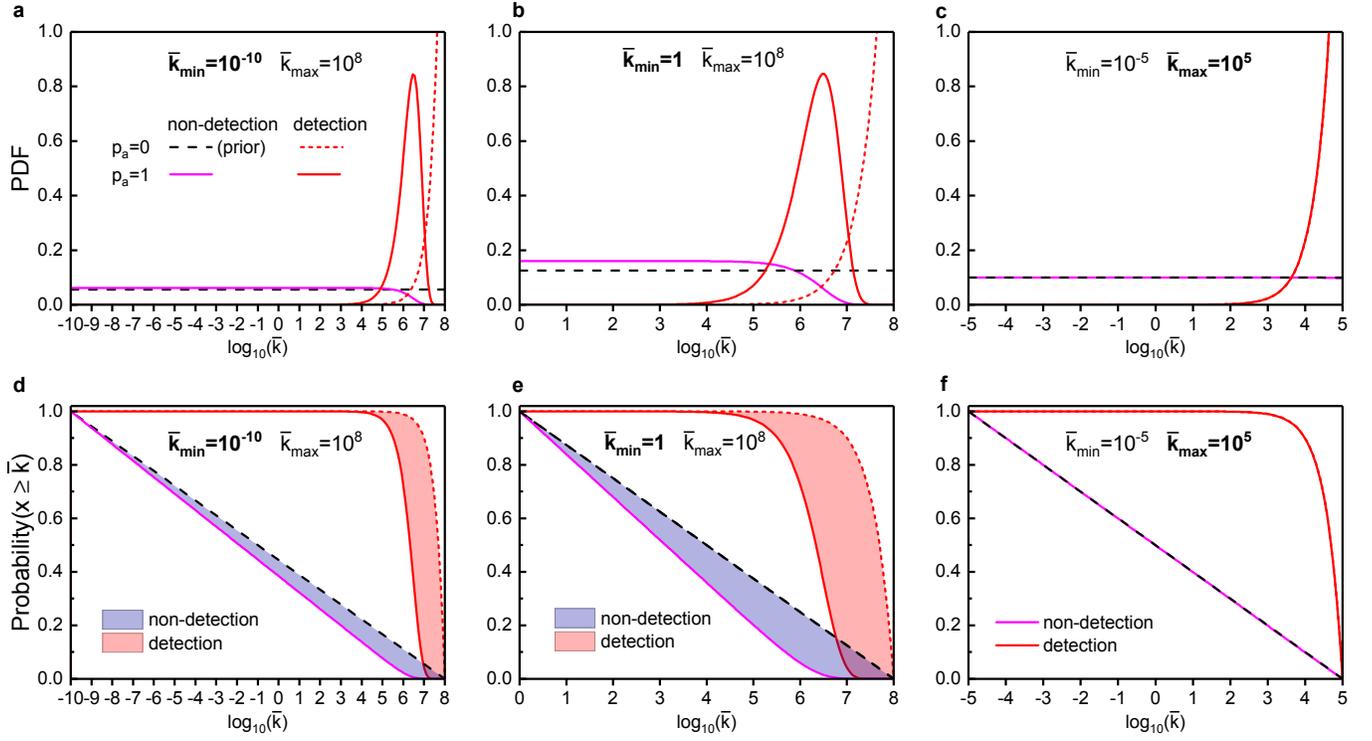

FIG. S8. PDFs (upper row) and CCDFs (bottom row) resulting from a survey within $R = 100$ ly calculated for different values of $\bar{k}_{\min}$ and $\bar{k}_{\max}$. The posteriors are calculated starting from a noninformative (log-uniform) prior. Solid lines refer to posteriors calculated for $p_a = 1$, while the black dashed lines and the red short dashed lines refer to the case $p_a = 0$. In the case of non-detection, the posteriors obtained in the limit $p_a = 0$ conicide with the prior. In **c** and **f** the posteriors for $p_a = 0$ and $p_a = 1$ practically coincide because $\bar{k}_{\max} \pi(R)$ is smaller than unity.



\end{document}